%% file: main.tex
\documentclass[10pt,conference]{IEEEtran}
\IEEEoverridecommandlockouts
\input{preamble}

\def\paperdoi{}% e.g. 10.1109/APSEC.2026.XXXXX
\makeatletter
\def\ps@IEEEtitlepagestyle{%
  \def\@oddhead{}\def\@evenhead{}%
  \def\@oddfoot{\copyrightnotice}\def\@evenfoot{\copyrightnotice}%
}
\newcommand{\copyrightnotice}{%
  \begin{minipage}[b]{\textwidth}\scriptsize\centering
  Accepted for publication in the Proceedings of the 33rd Asia-Pacific Software
  Engineering Conference (APSEC 2026), Early Research Achievements Track.%
  \ifx\paperdoi\@empty\else\ DOI: \href{https://doi.org/\paperdoi}{\paperdoi}\fi\\
  \copyright\,2026 IEEE. Personal use of this material is permitted. Permission
  from IEEE must be obtained for all other uses, in any current or future media,
  including reprinting/republishing this material for advertising or promotional
  purposes, creating new collective works, for resale or redistribution to
  servers or lists, or reuse of any copyrighted component of this work in other
  works.
  \end{minipage}%
}
\makeatother
\begin{document}

\title{Early Adoption, Use, and Maintenance of the Agent Package Manager in Open-Source Software}

\input{author_info}

\maketitle

\input{sections/abstract}
\input{sections/introduction}
\input{sections/background}
\input{sections/methodology}
\input{sections/rq1}
\input{sections/rq2}

\input{sections/rq3}
\input{sections/discussion}
\input{sections/threats}
\input{sections/conclusion}

\section*{Data Availability Statement}
Our replication package is available online~\cite{mind2026replication}.

\section*{Acknowledgments}
We gratefully acknowledge the financial support of: (1) JSPS for the KAKENHI grants (JP24K02921, JP25K03100, JP25K22845, JP26H02500); (2) Japan Science and Technology Agency (JST) as part of Adopting Sustainable Partnerships for Innovative Research Ecosystem (ASPIRE), Grant Number JPMJAP2415, and (3) the Inamori Research Institute for Science for supporting Yasutaka Kamei via the InaRIS Fellowship.

\bibliographystyle{IEEEtran}
\bibliography{references}

\end{document}

%% file: preamble.tex
\usepackage{cite}
\usepackage{amsmath,amssymb,amsfonts}
\usepackage{algorithmic}
\usepackage{graphicx}
\usepackage{textcomp}
\usepackage{xcolor}
\usepackage{booktabs}
\usepackage{hyperref}
\usepackage{multirow}
\usepackage{array}
\usepackage{url}
\usepackage{listings}
\usepackage{subcaption}

\usepackage{fontawesome5}
\usepackage{xspace}

\newcommand{\hand}[1]{%
  \noindent\faHandPointRight[regular]\xspace\textbf{\textit{#1}}%
}

\newcommand{\countobservations}{
    \def \countobservations{1}
}
\newcounter{observation}
\newenvironment{observation}[1]
{   \refstepcounter{observation}
    \noindent \textbf{Observation~\theobservation: \textbf{#1}}
}

\newcommand{\countimplications}{
    \def \countimplications{1}
}

\newcounter{implication}
\countobservations
\countimplications

\def\BibTeX{{\rm B\kern-.05em{\sc i\kern-.025em b}\kern-.08em
    T\kern-.1667em\lower.7ex\hbox{E}\kern-.125emX}}

\newcommand{\RQone}{(Configuration) How do developers configure APM to manage agent resources?}
\newcommand{\RQtwo}{(Lifecycle) What characterizes the lifecycle of APM in software repositories?}
\newcommand{\RQthree}{(APM Support) To what extent does APM automate agent-resource management?}

\lstdefinelanguage{yaml}{
  keywords={true,false,null,y,n},
  keywordstyle=\color{blue},
  basicstyle=\ttfamily\footnotesize,
  sensitive=false,
  comment=[l]{\#},
  commentstyle=\color{gray},
  morestring=[b]",
  morestring=[b]',
}

%% file: author_info.tex
\author{
    \IEEEauthorblockN{
        Napasorn Lapprakobkit\IEEEauthorrefmark{1}\IEEEauthorrefmark{2},
        Haruka Tokumasu\IEEEauthorrefmark{1}\IEEEauthorrefmark{3},
        Tao Xiao\IEEEauthorrefmark{3},
        Masanari Kondo\IEEEauthorrefmark{3}, \\
        Morakot Choetkiertikul\IEEEauthorrefmark{2},
        Chaiyong Ragkhitwetsagul\IEEEauthorrefmark{4},
        Yasutaka Kamei\IEEEauthorrefmark{3}
        \thanks{\IEEEauthorrefmark{1}These authors contributed equally to this work.}
    }
    \IEEEauthorblockA{
        \IEEEauthorrefmark{2}\textit{Faculty of Information and Communication Technology, Mahidol University}, Nakhon Pathom, Thailand \\
        napasorn.lar@student.mahidol.ac.th, morakot.cho@mahidol.ac.th
    }
    \IEEEauthorblockA{
        \IEEEauthorrefmark{3}\textit{Kyushu University}, Fukuoka, Japan \\
        tokumasu@posl.ait.kyushu-u.ac.jp, \{xiao, kondo, kamei\}@ait.kyushu-u.ac.jp
    }
    \IEEEauthorblockA{
        \IEEEauthorrefmark{4}\textit{Chulalongkorn University}, Bangkok, Thailand, chaiyong@cp.eng.chula.ac.th
    }
}

%% file: sections/abstract.tex
\begin{abstract}

The Agent Package Manager (APM) is an emerging dependency management system for AI agent resources, enabling developers to package, distribute, and reuse agent configurations through declarative manifests. Despite growing interest in reusable agent resources, little is known about how package-based management of these resources is adopted in practice. We present the first empirical study of APM, analyzing 126 GitHub repositories. We examine how developers configure APM, how it is introduced into existing projects, and how it interacts with manually maintained agent resources. Our findings suggest that APM primarily complements rather than replaces existing configurations, resulting in hybrid ecosystems where package-managed and manually maintained resources evolve together. These early observations provide the first empirical characterization of package-based management for AI agent resources and establish a foundation for future research on emerging agent software ecosystems.
\end{abstract}

\begin{IEEEkeywords}
Agent Package Manager, empirical study, package management
\end{IEEEkeywords}

%% file: sections/introduction.tex
\section{Introduction}
AI agents have become an important part of software development~\cite{agent4se_survey,se_agents_survey}, as they can automate various software engineering tasks, including code generation~\cite{yang2024sweagent}, software testing~\cite{Mundler2024SWT}, and code review~\cite{Tang2024CodeAgentAC}.
To fully leverage the capabilities of AI agents, developers must configure them appropriately, for example, by constructing suitable agent harnesses and selecting the skills required for the target tasks
~\cite{galster2026harnessengineering, xia2026harnessingagentskills}.
In team-based software development, these configurations must also be shared among developers and continuously maintained as the project evolves.

To support the team-level sharing and maintenance of AI agent configurations, Microsoft introduced the Agent Package Manager (APM)~\cite{microsoft_apm}.
APM allows developers to configure AI agent resources (e.g., declare dependencies, select skills, and specify MCP server targets) in an \texttt{apm.yml} manifest. It automatically resolves and installs these dependencies, while recording the resolved versions in an \texttt{apm.lock.yaml} file.
This mirrors conventional package managers such as npm.

In this paper, we present the first empirical study of the adoption, usage, and maintenance of APM in open-source software. We investigate how developers adopt package-based management for AI agent resources and how APM use shapes the boundary between package-managed and manually maintained resources. Our analyses provide insights into the early adoption and usage patterns of APM. These insights can inform both the design of future agentic development tools and recommendations
for using APM to manage AI agent resources. We analyze 126 GitHub repositories that use APM to answer the following three research questions:

\begin{itemize}
    \item \textbf{RQ1}: \textbf{\emph{\RQone}}
    Configuring AI agent resources is a new software engineering task. This RQ aims to understand how developers use the APM manifest to express their configuration choices.

    \item \textbf{RQ2}: \textbf{\emph{\RQtwo}}
    APM can be introduced into a repository in two ways: by migrating existing local agent files into APM packages or by bundling APM alongside existing AI agent resources.
    Also, APM evolves over time as developers add, remove, and update configurations. This RQ investigates how APM is introduced and how it evolves over time.

    \item \textbf{RQ3}: \textbf{\emph{\RQthree}}
    The primary goal of APM is to manage the AI agent resources to support developers in sharing and maintaining them. This RQ investigates
    how much
    APM manages the agent resource ecosystem compared to manually maintained files.

\end{itemize}

%% file: sections/background.tex
\section{Background}
\label{sec:background}

\input{img/apm_yml_and_lock_figures}
\subsection{Package and AI Skills Management}
Package management is a crucial mechanism for managing dependencies and ensuring reproducibility, supported by mature tools such as npm for JavaScript, pip for Python, and Maven for Java. Prior studies examined established ecosystems to clarify their adoption, use, and dependency management: Decan et al.~\cite{decan2019dependencynetwork} studied the evolution of seven package ecosystems, and Kula et al.~\cite{kula2017npm} showed how npm dependency structures can become complex. These studies focus on traditional software libraries, whereas APM manages AI-agent resources such as skills and configurations.

To manage AI agent skills, developers can use tools such as \texttt{npx skills}~\cite{npxskills} and \texttt{gh skill}~\cite{ghskill}. However, these tools do not provide a unified package management mechanism for different types of agent resources, and they primarily target local development, offering limited support for sharing and maintaining agent configurations across team members. APM addresses these limitations by providing a structured mechanism for declaring dependencies on reusable agent resources and managing their configurations at the project level.

We thus present an initial study of APM adoption and use.

\subsection{Agent Package Manager (APM)}
APM supports project-level management and sharing of agent configurations.
Once a developer creates a manifest file (\texttt{apm.yml}), they can run \texttt{apm install} to resolve the declared dependencies and deploy the required resources to agent-specific directories, such as
\texttt{.github/skills/}. The corresponding lockfile (\texttt{apm.lock.yaml}) records the resolved package versions and the resources deployed from each package.

Figure~\ref{fig:manifests}
illustrates an example manifest and its lockfile. The manifest contains several sections: the \texttt{metadata} block provides descriptive information, while \texttt{APM dependencies} and \texttt{MCP servers} list the dependencies to be fetched and distributed to the agents.
Other fields (e.g., targets) are optional and often resolved automatically.
For each dependency, the lockfile records the deployment destination (\texttt{deployed\_files}), the version pin, and, when possible, the content integrity hash.

Note that APM need not manage all agent resources in a repository. Alongside APM-managed files, developers can still maintain agent configuration files manually, such as agent context files (e.g., \texttt{AGENTS.md}) and local skills.

%% file: img/apm_yml_and_lock_figures.tex
\begin{figure}[t]
\centering

\begin{subfigure}[t]{0.8\linewidth}
\centering
\includegraphics[width=\linewidth]{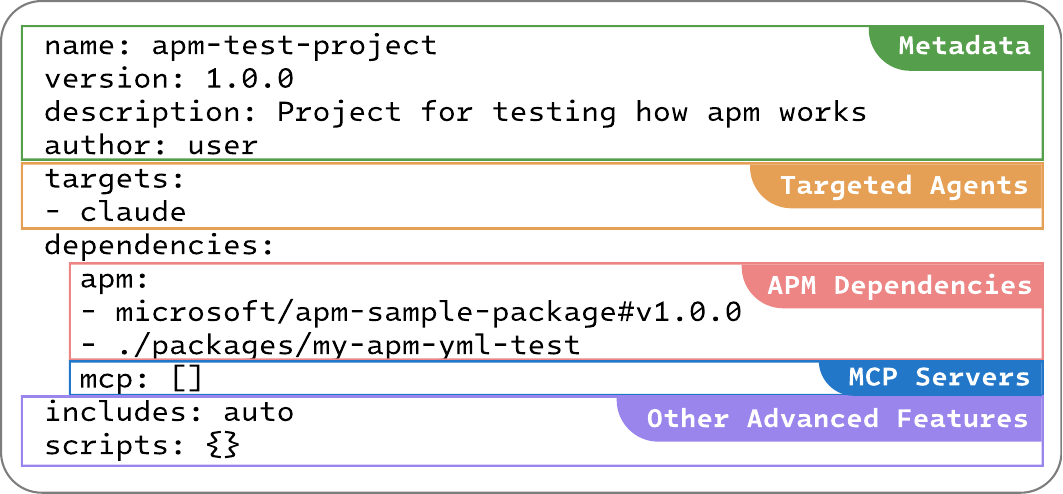}
\caption{\texttt{apm.yml} structure example}
\label{fig:apm_yml_structure}
\end{subfigure}

\begin{subfigure}[t]{0.8\linewidth}
\centering
\includegraphics[width=\linewidth]{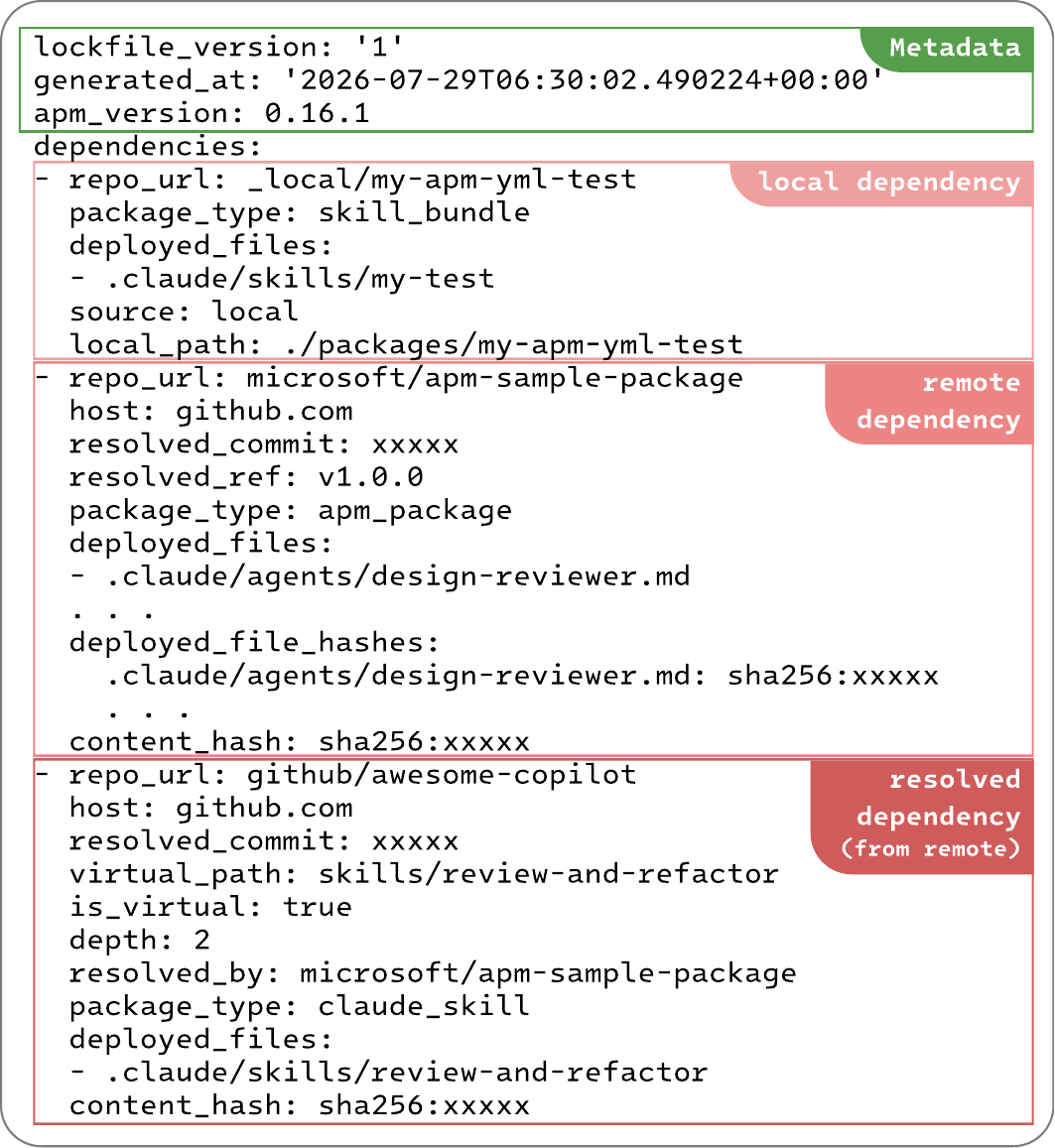}
\caption{\texttt{apm.lock.yaml} structure example}
\label{fig:apm_lock_structure}
\end{subfigure}

\caption{APM manifest files structure.}
\label{fig:manifests}

\end{figure}

%% file: sections/methodology.tex
\input{img/Highlevel_overview_diagram}

\section{Data Collection}
\label{sec:methodology}

Figure~\ref{highlevel_overview} illustrates the overview of our study methodology. We explain each step below.

\subsection{Data Collection Pipeline}

Our pipeline comprises four sequential stages.

\noindent{}
\textbf{Stage 1: Repository discovery:} We used the GitHub Code Search API to find all public files named \texttt{apm.yml} and \texttt{apm.lock.yaml}, de-duplicating the results
by \texttt{owner/repo:filepath}. We retained only repositories containing both files, to
distinguish genuine APM users from repositories that happen to have an \texttt{apm.yml} for an unrelated purpose (e.g., Elastic APM monitoring config).

\noindent{}
\textbf{Stage 2: Manual inspection:}
For the candidate repositories, the first and second authors manually determined whether APM was actually used rather than merely used as a template or example.
To measure inter-rater reliability, two authors independently reviewed a random sample of 20 repositories (Cohen's $\kappa = 0.898$ ~\cite{Cohen1960ACO}, above the 0.80 threshold for ``almost perfect agreement''~\cite{landis1977}); the first author then inspected the remaining repositories, with ambiguous cases resolved through discussion between two authors. This yielded a final dataset of 126 included repositories, frozen on 3 June 2026; all subsequent data collection was performed on these repositories.

\noindent{}
\textbf{Stage 3: Commit collection:} For each repository, we fetched all commits touching \texttt{apm.yml} or \texttt{apm.lock.yaml} using the GitHub Commits API, and retrieved the full diff of each from the Commit Details API to capture co-changed files. If API retrieval failed, we cloned the repository locally.

\noindent{}
\textbf{Stage 4: Additional collection:}
For all repositories, we additionally collected: (a) the file tree at the commit immediately before the first \texttt{apm.yml} commit, (b) GitHub issues created after APM adoption, and (c) the current \texttt{apm.lock.yaml} and complete file tree for each repository.

\subsection{Data Extraction}
For each repository, we collected the commit history of \texttt{apm.yml} and extracted the manifest content from each file revision to identify the final manifest state.
Each extracted manifest was parsed as a YAML document and processed using a schema-based field extraction approach. Specifically, we extracted values from predefined YAML keys representing major components of an APM manifest, including package metadata (\texttt{name}, \texttt{version}, and \texttt{type}), dependency declarations (\texttt{dependencies.apm}), deployment configuration (\texttt{target}/\texttt{targets}), and optional capabilities such as MCP servers and scripts (\texttt{dependencies.mcp} and \texttt{scripts}). The selected fields were based on their roles within the manifest and their relevance to our research questions.

%% file: img/Highlevel_overview_diagram.tex
\begin{figure}[t]
\centering
    \includegraphics[width=0.45\textwidth]{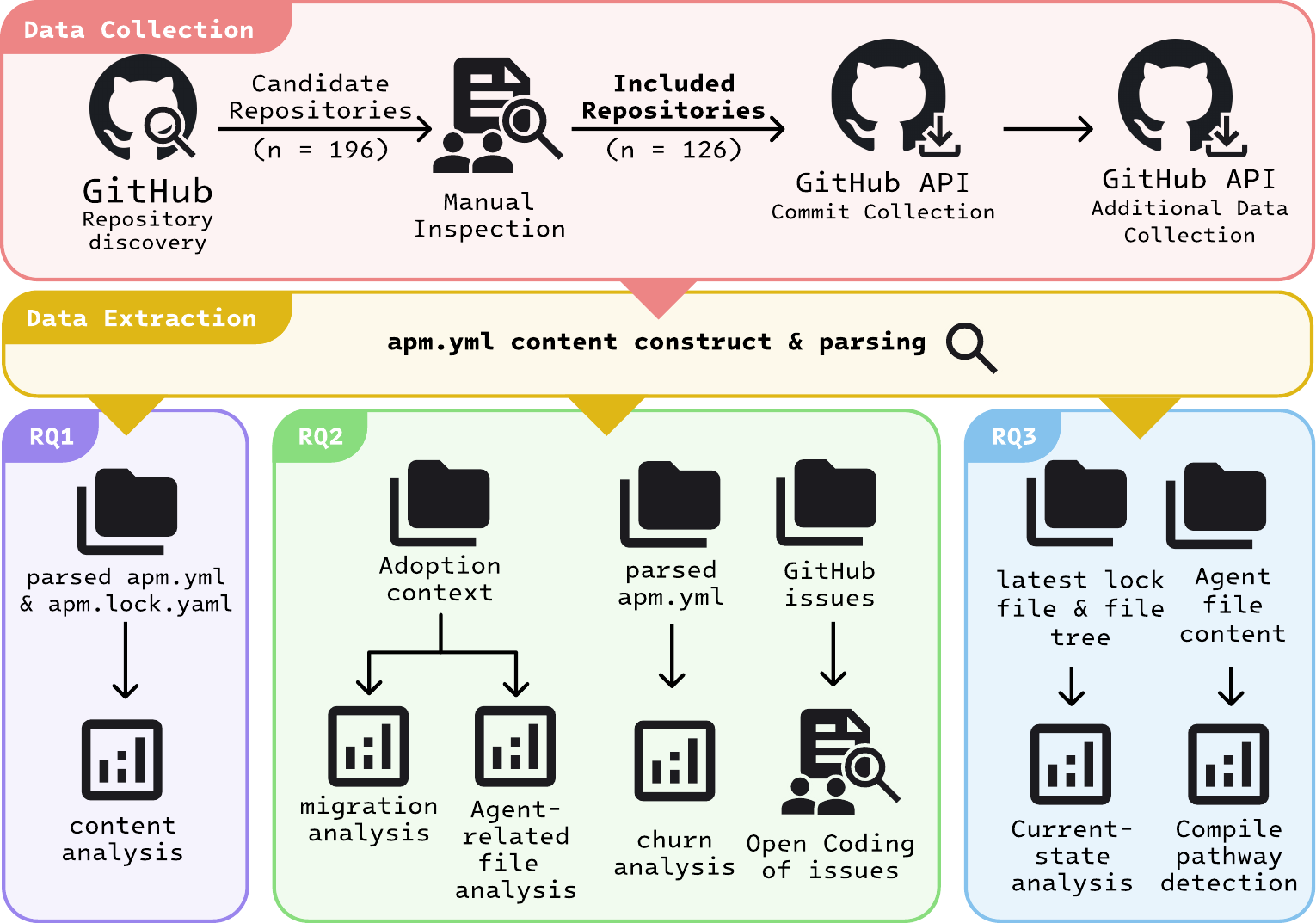}
    \captionof{figure}{Overview of the study methodology.}
    \label{highlevel_overview}
\end{figure}

%% file: sections/rq1.tex
\section{RQ1: \RQone}
\label{sec:rq1}

\subsection{Approach}

The content analysis operates on the \textbf{192} parsed \texttt{apm.yml} manifest files from all \textbf{126} repositories. We also analyze the corresponding \texttt{apm.lock.yaml} files and compare them with the manifests. We extract deployment paths from the lockfile to identify the agents recorded by APM.

\subsection{Results}
\observation{apm.yml manifests primarily serve as simple dependency declarations.}
Table~\ref{tab:configuration}.\subref{tab:feature-adoption} shows the manifest features adopted in the 126 repositories.
APM dependencies appear in 116 repositories (92.1\%), while the next most common feature, target agents, appears in only 51 (40.5\%).
These results indicate that dependency declarations are the central component of APM configurations, while other features are less commonly used.
Table~\ref{tab:configuration}.\subref{tab:dependency-distribution} shows the distribution of the number of APM dependencies per manifest. Among the 127 of 192 manifest files that declared at least one APM dependency, 82 (64.6\%) declare only one to three dependencies. These results suggest that dependency declarations are typically small in scale.

The lockfile analysis further shows that declared targets do not always reflect actual deployments. Table~\ref{tab:target-discrepancy} compares targets specified in the manifest with agents identified from the lockfile. For example, Claude is declared as a target in 33 repositories but appears in the deployment paths of 61 repositories.
These results show that APM manifests can remain simple even when packages are deployed across multiple agents.

\input{table/tab_features}
\input{table/tab_target}

\vspace{6pt}
\noindent\fbox{\parbox{0.97\columnwidth}{%
\textbf{Summary (RQ1).}
Repositories primarily adopt APM for dependency management. Their \texttt{apm.yml} manifests generally contain only a small number of dependencies, while other manifest features are less commonly used.
The lockfile also shows that actual deployments often involve more agents than explicitly declared in the manifest.
}}
\vspace{6pt}

%% file: table/tab_features.tex
\begin{table}[t]
\centering
\caption{Configuration of \texttt{apm.yml} manifests}
\label{tab:configuration}
\scriptsize
\setlength{\tabcolsep}{2pt}

\begin{subtable}[t]{0.55\columnwidth}
\centering
\caption{Feature adoption}
\label{tab:feature-adoption}
\begin{tabular}{@{}lr@{}}
\toprule
\textbf{Feature} & \textbf{Repos.(\%)} \\
\midrule
APM dependencies  & 116 (92.1\%) \\
Target agent  & 51 (40.5\%) \\
MCP servers  & 14 (11.1\%) \\
Compilation key  & 4 (3.2\%) \\
devDependencies  & 3 (2.4\%) \\
Scripts  & 7 (5.6\%) \\
Primitives key  & 1 (0.8\%) \\
Marketplace key  & 1 (0.8\%) \\
\bottomrule
\end{tabular}
\end{subtable}
\hfill
\begin{subtable}[t]{0.41\columnwidth}
\centering
\caption{Dep. per manifest}
\label{tab:dependency-distribution}
\begin{tabular}{@{}lr@{}}
\toprule
\textbf{Dependencies} & \textbf{Instances (\%)} \\
\midrule
    0         & 65 (33.9\%) \\
    1         & 51 (26.6\%) \\
    2         & 20 (10.4\%) \\
    3         & 11 (5.7\%)  \\
    4         & 9 (4.7\%)   \\
    5         & 7 (3.6\%)   \\
    6         & 6 (3.1\%)   \\
    7 or more & 23 (12.0\%) \\
\bottomrule
\end{tabular}
\end{subtable}

\end{table}

%% file: table/tab_target.tex
\begin{table}[t]
\centering
\caption{Comparison of target declarations in \texttt{apm.yml} and deployed agents identified from \texttt{apm.lock.yaml}.}
\label{tab:target-discrepancy}
\begin{tabular}{lrr}
\toprule
Target & Declared & Deployed \\
\midrule
Claude  & 33 & 61 \\
Copilot & 25 & 57 \\
Codex   & 11 &  9 \\
Cursor  &  5 &  6 \\
Gemini  &  4 &  4 \\
\bottomrule
\end{tabular}
\end{table}

%% file: sections/rq2.tex
\section{RQ2: \RQtwo}
\label{sec:rq2}

\subsection{Approach}

The analysis combines three views of the APM lifecycle across all 126 repositories.
First, we analyze the repository state before and after APM adoption using the file tree before the first \texttt{apm.yml} commit, and the adoption commit itself. We identify agent-related files and use \texttt{apm.lock.yaml} to distinguish files managed by APM from files added manually alongside APM. Changes to these files are tracked to distinguish migration from bundling.
Second,
we analyze the lifecycle of agent-related files across three repository
states: the file tree before APM adoption, the adoption commit, and the current file tree. These snapshots are used to examine changes in the presence and number of agent-related files before and after APM adoption.
Finally, manifest maintenance is measured using commit churn. Files with
$\leq$3 commits are classified as \emph{low-churn},
others as \emph{high-churn}. High-churn manifests are then analysed by their change types.

To complement the quantitative findings, we also performed an open coding of GitHub issues to identify their discussion topics and determine whether they are related to manifest files.
We coded 48 APM-mentioning issues along two independent dimensions (Issue Type and APM Role) using an open coding procedure~\cite{wang2023conversationsturnworktaxonomy}.
Codes were developed through repeated examination and discussion rather than being defined in advance.
Two researchers jointly open-coded a random sample of 20 issues to develop and stabilise the codebook. A further 10 issues were then coded independently by both researchers to measure inter-rater reliability; Cohen's $\kappa = 0.846$ for Issue Type and $\kappa = 0.853$ for APM Role~\cite{Cohen1960ACO}, both above the 0.80 threshold for ``almost perfect agreement''~\cite{landis1977}. The first author then applied the finalised codes to all remaining issues. Two issues were excluded as irrelevant,
 leaving N\,=\,46 coded issues.

\subsection{Results}

\observation{APM adoption is predominantly a bundling decision, not a migration.}
Among all 126 repositories, 99 repositories (78.6\%) did not modify any local agent files.
Thus, more than half of the repositories adopted APM without replacing existing local agent files.
In contrast, 11 repositories (8.7\%) removed existing local agent files and replaced them with APM packages.
When migration does occur, three strategies are observed: (1)
moving files into a personal APM package, (2)
replacing per-agent config files with cross-agent APM instruction packages, and (3)
migrating from project-specific instruction files to APM-managed resources.
These patterns indicate that APM supports migrating both reusable and project-specific agent content, although such migrations were observed in only a small proportion of repositories.

\observation{Companion agent files are introduced with APM and persist after adoption.}
Table~\ref{tab:companion-files} reports the number of repositories in which each companion agent file appeared before APM adoption, in the APM creation commit, and in the current file tree.
\texttt{SKILL.md} showed the largest increase, from 24 repositories before adoption to 86 at the time of creation. For each tracked file, the Ongoing count exceeded the Creation count, indicating that repositories generally continued to contain these agent-related files after APM adoption.

\observation{APM manifests are rarely modified after creation, and later changes primarily add dependencies rather than revise versions or structure.}
83.9\% of \texttt{apm.yml} instances had at most 3 commits. This limited churn contrasts with the expectation that configuration-as-code files evolve alongside the codebase.
Among the 31 high-churn manifests with more than three commits, 87.1\% included dependency additions, whereas 51.6\% included version or reference changes.
This indicates that \texttt{apm.yml} is primarily used to declare dependencies rather than maintained as a continuously evolving specification.

The results of the open coding
are consistent with this pattern. Of 46 issues, 14 (30.43\%) focus on source files (e.g., local dependency files) rather than the manifest itself. Additionally, although 4 out of 8 bug reports attribute problems to APM, fixes suggest workarounds rather than changes to the manifest. This is consistent with the low observed churn of \texttt{apm.yml}.

\vspace{6pt}
\noindent\fbox{\begin{minipage}{\dimexpr\columnwidth-2\fboxsep-2\fboxrule\relax}
\smallskip
\textbf{Summary (RQ2).}
APM is usually introduced by adding package management alongside existing agent files rather than replacing them. The agent-related files created during adoption remain in use throughout later development, while the manifest itself changes little, mainly by adding dependencies instead of updating versions.
\smallskip
\end{minipage}}

\input{table/companion_files}

%% file: table/companion_files.tex
\begin{table}[t]
\centering
\caption{Agent companion files across the APM lifecycle.}
\label{tab:companion-files}
\begin{tabular}{@{}lrrr@{}}
\toprule
\textbf{Basename} & \textbf{Before} & \textbf{Creation} & \textbf{Ongoing} \\
\midrule
\texttt{SKILL.md}      & 24 &  86 & 102 \\
\texttt{AGENTS.md}     & 28 &  45 &  66 \\
\texttt{CLAUDE.md}     & 32 &  41 &  42 \\
\texttt{copilot-instructions.md}  & 22 & 26 & 30 \\
\texttt{GEMINI.md}     &  2 &  10 &  14 \\
\texttt{.mcp.json}     &  5 &  13 &  21 \\
\bottomrule
\end{tabular}
\end{table}

%% file: sections/rq3.tex
\section{RQ3: \RQthree}
\label{sec:rq3}

\subsection{Approach}
The analysis consists of two complementary parts.
\textit{Current-state analysis} examines the coexistence of APM-managed and manually maintained agent files. The analysis uses the current lockfile
and the complete file tree of all 126 repositories. Files listed under paths in the
file deployment
fields of the lockfile are classified as APM-managed, while all remaining agent-related files are classified as manually maintained.

\textit{Automated-resource detection} examines the two mechanisms through which APM introduces agent resources. The \texttt{apm install} command deploys resources from declared APM packages into a repository, including skill files.
The \texttt{apm compile} command generates agent-specific configuration files
(such as \texttt{AGENTS.md}, etc.),
from resources managed by APM. Installed files are identified from \texttt{apm.lock.yaml}, while compiled files are identified using content markers such as \texttt{Generated by APM CLI}.

\subsection{Results}

\observation{APM-managed and manually maintained agent files coexist in most repositories.}
Of the 126 repositories, 78 (61.9\%) contained both APM-managed and manually maintained agent files. In comparison, 14 repositories (11.1\%) contained only APM-managed files, 25 (19.8\%) contained only manually maintained files, and 9 (7.1\%) contained neither. The median number of files per repository was 13 for APM-managed files and 3 for manually maintained files.
This indicates that manual maintenance of agent-related files remains common after APM adoption.

\observation{APM automation is concentrated on skill deployment, while automated generation of agent configurations is less common.}
The \texttt{apm install} command automatically deployed skill files in 84 of the 126 repositories (66.7\%). Automated generation of agent configuration files through \texttt{apm compile} was identified in 18 repositories (14.3\%). The compiled outputs included \texttt{AGENTS.md}, \texttt{CLAUDE.md}, and \texttt{GEMINI.md}. Among the 249 tracked files, 58 (23.3\%) were generated by APM, while 191 (76.7\%) were authored manually. The results indicate that APM automates skill deployment more widely than the generation of configurations for different agents.

\vspace{6pt}
\noindent\fbox{\begin{minipage}{\dimexpr\columnwidth-2\fboxsep-2\fboxrule\relax}
\smallskip
\textbf{Summary (RQ3).}
APM automates selected parts of agent-resource management by deploying skills and generating agent configuration files. However, APM-managed and manually maintained files commonly coexist.

\smallskip
\end{minipage}}
\vspace{6pt}

%% file: sections/discussion.tex
\section{Discussion}
\label{sec:discussion}

In this section, we discuss the implications (see \hand{}icon) of our findings.

\textbf{Broader use of APM capabilities requires clearer onboarding and communication of their practical value.}
Observation 1 shows that APM usage is largely limited to dependency management, while other capabilities are rarely adopted. This result indicates that APM capabilities are used unevenly across repositories. The limited use of broader capabilities may reflect unclear use cases or insufficient communication of their practical benefits.

\hand{For APM developers,} \textit{clearer use cases, practical examples, and guided setup should be provided to help users understand when and how each capability can be used.}

\textbf{Manifest declarations should be distinguished from the actual management state of agent-related files.}
APM is designed to manage agent-related resources specified through the manifest rather than the entire agent environment. Observation 1 shows that target agents
are often unspecified,
limiting the ability of the manifest to reveal where agent-related files are deployed. Observations 2 and 5 show that APM-managed files coexist with manually maintained files after adoption; 76.7\% of the tracked agent-configuration files were manually maintained.
This suggests that the manifest alone may give an incomplete picture, and that the lockfile is a more reliable record of which files APM manages and deploys.

\hand{For APM users,} \textit{the lockfile should be inspected after installation and updates to confirm that the intended files are managed and deployed to the expected locations. Comparing the lockfile with the repository file tree can also reveal unintended changes.}

\hand{For researchers,} \textit{the management status of agent-related files should not be inferred solely from manifest contents or file paths. The lockfile should be combined with repository file trees and change histories to identify APM-managed files, measure the actual scope of APM management, and examine how management responsibility changes over time.}

%% file: sections/threats.tex
\section{Threats to Validity}
\label{sec:threats}
\noindent{}
\textbf{Construct Validity.}
Manual repository selection in our data collection may introduce human bias or errors.
We mitigate this threat by applying consistent inclusion criteria
and resolving ambiguous cases through discussion between two authors.

\smallskip{}
\noindent{}
\textbf{Internal Validity.}
Our experimental script may include errors that could affect the accuracy of our results.
Hence, we disclosed our scripts to enable replication and verification of our results.

\smallskip{}
\noindent{}
\textbf{External Validity.}
Our dataset consists of 126 public GitHub repositories. Since APM is still in its early stages of adoption, the number of repositories using it was limited.
Hence, our findings might be an early characterization of APM adoption and usage.

%% file: sections/conclusion.tex
\section{Conclusion}
\label{sec:conclusion}
We presented an early empirical study of APM use in 126 open-source repositories.
Our findings suggest that APM is currently used primarily for lightweight dependency declaration and skill deployment, and it usually complements rather than replaces existing agent resources. Consequently, most studied repositories retain a hybrid management model in which APM-managed and manually maintained files coexist.
Specifically, APM primarily manages skill deployment, while agent configuration files are still maintained manually.

\textbf{Future work.}
To improve the generalizability of our findings, we plan to extend our dataset to include a larger number of repositories.
Since APM is still a relatively new tool, we also plan to conduct longitudinal studies to examine how its adoption and management practices evolve over time.
In addition, to investigate the impact of APM adoption on developer productivity and development activities, we plan to examine whether adopting APM is associated with changes in the frequency and characteristics of commits, pull requests, and issues.

%% file: references.bib
@misc{microsoft_apm,
  author       = {{Microsoft}},
  title        = {Agent Package Manager ({APM})},
  year         = {2025},
  howpublished = {\url{https://github.com/microsoft/apm}}
}

@misc{npxskills,
  author       = {{Vercel Labs}},
  title        = {skills: The {CLI} for the Open Agent Skills Ecosystem},
  year         = {2026},
  howpublished = {\url{https://www.npmjs.com/package/skills}}
}

@misc{ghskill,
  author       = {{GitHub}},
  title        = {{GitHub CLI}: gh skill},
  year         = {2026},
  howpublished = {\url{https://cli.github.com/manual/gh_skill}}
}

@misc{mind2026replication,
  author       = {Lapprakobkit, Napasorn and others},
  title        = {Replication Package},
  year         = {2026},
  howpublished = {\url{https://doi.org/10.5281/zenodo.21833769}}
}

@article{Cohen1960ACO,
  author  = {Cohen, Jacob},
  title   = {A Coefficient of Agreement for Nominal Scales},
  journal = {Educ. Psychol. Meas.},
  year    = {1960},
  volume  = {20},
  number  = {1},
  pages   = {37--46}
}

@article{landis1977,
  author  = {Landis, J. Richard and others},
  title   = {The Measurement of Observer Agreement for Categorical Data},
  journal = {Biometrics},
  year    = {1977},
  volume  = {33},
  number  = {1},
  pages   = {159--174}
}

@article{wang2023conversationsturnworktaxonomy,
  author  = {Wang, Dong and others},
  title   = {When Conversations Turn Into Work: A Taxonomy of Converted Discussions and Issues in {GitHub}},
  journal = {Empir. Softw. Eng.},
  year    = {2023},
  volume  = {28},
  number  = {6},
  doi     = {10.1007/s10664-023-10366-z}
}

@article{agent4se_survey,
  author  = {Liu, Junwei and others},
  title   = {Large Language Model-Based Agents for Software Engineering: A Survey},
  journal = {ACM Trans. Softw. Eng. Methodol.},
  year    = {2026},
  doi     = {10.1145/3796507}
}

@article{se_agents_survey,
  author  = {Wang, Yanlin and others},
  title   = {Agents in Software Engineering: Survey, Landscape, and Vision},
  journal = {Autom. Softw. Eng.},
  year    = {2025},
  volume  = {32},
  number  = {2},
  doi     = {10.1007/s10515-025-00544-2}
}

@inproceedings{yang2024sweagent,
  author    = {Yang, John and others},
  title     = {{SWE-agent}: Agent-Computer Interfaces Enable Automated Software Engineering},
  booktitle = {Proc.\ NeurIPS},
  year      = {2024}
}

@inproceedings{Mundler2024SWT,
  author    = {M\"{u}ndler, Niels and others},
  title     = {{SWT-Bench}: Testing and Validating Real-World Bug-Fixes with Code Agents},
  booktitle = {Proc.\ NeurIPS},
  year      = {2024}
}

@inproceedings{Tang2024CodeAgentAC,
  author    = {Tang, Xunzhu and others},
  title     = {{CodeAgent}: Autonomous Communicative Agents for Code Review},
  booktitle = {Proc.\ EMNLP},
  year      = {2024}
}

@article{galster2026harnessengineering,
  author  = {Galster, Matthias and others},
  title   = {Harness Engineering for Agentic {AI} Coding Tools: An Exploratory Study},
  journal = {arXiv preprint arXiv:2602.14690},
  year    = {2026}
}

@misc{xia2026harnessingagentskills,
  author = {Xia, Boming and others},
  title  = {Harnessing Agent Skills: Architectural Patterns and a Reference Architecture for Skill-Mediated {LLM} Agents},
  year   = {2026},
  note   = {{SSRN} preprint},
  doi    = {10.2139/ssrn.6871959}
}

@article{decan2019dependencynetwork,
  author  = {Decan, Alexandre and others},
  title   = {An Empirical Comparison of Dependency Network Evolution in Seven Software Packaging Ecosystems},
  journal = {Empir. Softw. Eng.},
  year    = {2019},
  volume  = {24},
  number  = {1},
  pages   = {381--416},
  doi     = {10.1007/s10664-017-9589-y}
}

@article{kula2017npm,
  author  = {Kula, Raula Gaikovina and others},
  title   = {On the Impact of Micro-Packages: An Empirical Study of the {npm} {JavaScript} Ecosystem},
  journal = {arXiv preprint arXiv:1709.04638},
  year    = {2017}
}
